\documentstyle[12pt]{article}
\begin{document}

%\begin{titlepage}
%\thispagestyle{empty}

\begin{center}
{\Large\bf
Effective action in general chiral superfield model}
\end{center}

\begin{center}
{\bf
A.Yu.Petrov%\footnote{e-mail: petrov@tspu.edu.ru}
}\\
\footnotesize{
{\it
Department of Theoretical Physics,
Tomsk State Pedagogical University\\
Tomsk 634041, Russia}\\
}
\end{center}

\vspace*{.2cm}

\normalsize

%\section{Introduction}
According to the superstring theory the low-energy elementary particle
models
contain as ingredient the multiplets of chiral and antichiral superfields action
of which is given in terms of k\"{a}hlerian effective potential
$K(\bar{\Phi},\Phi)$ and chiral $W(\Phi)$ and antichiral $\bar{W}(\bar{\Phi})$
potentials. These potentials are found in explicit and closed form
within string perturbation theory (see f.e. \cite{GSW}).
Phenomenological aspects of such models have been studied in recent papers
\cite{Cv}. In quantum theory one can expect an appearance of quantum corrections
to the potentials $K(\bar{\Phi},\Phi)$ and $W(\Phi)$. As a result we face a
problem of calculating effective action in models with arbitrary functions
$K(\bar{\Phi},\Phi)$ and $W(\Phi)$.

The remarkable features of the massless theories with $N=1$ chiral
superfields are the possibilities of obtaining the chiral quantum
corrections.  A few years ago West \cite{West2} pointed out that finite
two-loop chiral contribution to effective action really arises in
massless Wess-Zumino model (see also \cite{Buch3}).

In this talk we consider the general problem of calculating leading quantum
correction to chiral potential and k\"{a}hlerian potential in theory with
arbitrary potentials $K(\bar{\Phi},\Phi)$ and $W(\Phi)$, $\bar{W}(\bar{\Phi})$.
The remarkable result we obtain here
is that despite the theory under consideration is non-renormalizable at
arbitrary $K(\bar{\Phi},\Phi)$, $W(\Phi)$,
the lower (two-loop) chiral correction to effective action is always
finite.

%\section{Structure of effective action}

We consider $N=1$ supersymmetric field theory with action
\begin{equation}
\label{act}
S[\bar{\Phi},\Phi]=\int d^8 z K(\bar{\Phi},\Phi)+
(\int d^6 z W(\Phi)+h.c.)
\end{equation}
where $\Phi(z)$ and $\bar{\Phi}(z)$ are chiral and antichiral superfields
respectively. As well known, the real function $K(\bar{\Phi},\Phi)$
is called k\"{a}hlerian potential and holomorphic function $W(\Phi)$ is
called chiral potential \cite{BK0}. The partial cases of the theory (1) are
Wess-Zumino model with $K(\bar{\Phi},\Phi)=\Phi\bar{\Phi}$, $W(\Phi)\sim\Phi^3$
and $N=1$ supersymmetric four-dimensional sigma-model with $W(\Phi)=0$.
The action (1) is a most general one constructed from chiral and antichiral
superfields which does not contain the higher derivatives at component level.
Therefore it is natural to call the theory (1) a general chiral superfield
model.

Let $\Gamma [\bar{\Phi},\Phi]$ be effective action in the model (1).
Within a momentum expansion the effective action can be presented as a series
in supercovariant derivatives
$D_A=(\partial_a,D_{\alpha},\bar{D}_{\dot{\alpha}})$ in the form
\begin{eqnarray}
 \Gamma[\bar{\Phi},\Phi] &=& \int d^8z {\cal L}_{eff}
(\Phi,D_A\Phi,D_A D_B\Phi;\bar{\Phi},D_A\bar{\Phi},D_A D_B\bar{\Phi})
+\nonumber\\
&+&(\int d^6z {\cal L}^{(c)}_{eff}(\Phi) + h.c.) +\ldots
\end{eqnarray}
Here
${\cal L}_{eff}$ is called general effective lagrangian,
${\cal L}^{(c)}_{eff}$ is called chiral effective lagrangian. Both these
lagrangians are the series in supercovariant derivatives
of superfields and can be written as follows
\begin{eqnarray}
\label{ep}
{\cal L}_{eff}&=&K_{eff}(\bar{\Phi},\Phi)+\ldots
=K(\bar{\Phi},\Phi)+\sum_{n=1}^{\infty}K^{(n)}_{eff}(\bar{\Phi},\Phi)\nonumber\\
{\cal L}^{(c)}_{eff}&=&W_{eff}(\Phi)+\ldots=
W(\Phi)+\sum_{n=1}^{\infty}W^{(n)}_{eff}(\Phi)+\ldots
\end{eqnarray}
Here dots mean terms depending on covariant derivatives
of superfields
$\Phi,\bar{\Phi}$. Here $K_{eff}(\bar{\Phi},\Phi)$ is called
kahlerian effective potential, $W_{eff}(\Phi)$ is called chiral (or holomorphic)
effective potential,
$K^{(n)}_{eff}$ is a $n$-th correction to kahlerian potential
and $W^{(n)}_{eff}$ is a $n$-th correcton to chiral (holomorphic)
potential $W$.

To consider the effective lagrangians ${\cal L}_{eff}$ and
${\cal L}^{(c)}_{eff}$ we use path integral representation of
the effective action \cite{BK0,BO}
\begin{eqnarray}
\label{Green1}
 \exp(\frac{i}{\hbar}\Gamma[\bar{\Phi},\Phi]) &=&
 \int {\cal D} \phi {\cal D} \bar{\phi}
 \exp\big(
\frac{i}{\hbar}
S[\bar{\Phi}+\sqrt{\hbar}\bar{\phi},\Phi+\sqrt{\hbar}\phi]
-\nonumber\\&-&
(\int d^6 z
\frac{\delta\Gamma[\bar{\Phi},\Phi]}{\delta\Phi(z)}\phi(z)+h.c. )
\big)
\end{eqnarray}
Here $\Phi,\bar{\Phi}$ are the background superfields and $\phi,\bar{\phi}$
are the quantum ones.
The effective action can be written as
$\Gamma[\bar{\Phi},\Phi]=S[\bar{\Phi},\Phi]+\tilde{\Gamma}[\bar{\Phi},\Phi]$,
where $\tilde{\Gamma}[\bar{\Phi},\Phi]$ is a quantum correction.
Eq. (\ref{Green1}) allows to obtain $\tilde{\Gamma}[\bar{\Phi},\Phi]$ in form of
loop expansion
%\begin{equation}
$\label{Gamma}
 \tilde{\Gamma}[\bar{\Phi},\Phi] = \sum_{n=1}^{\infty}\hbar^n
\Gamma^{(n)} [\bar{\Phi},\Phi]$
%\end{equation}
and hence, to get loop expansion for the effective lagrangians
${\cal L}_{eff}$ and ${\cal L}^{(c)}_{eff}$.

To find loop corrections $\Gamma^{(n)} [\bar{\Phi},\Phi]$ in explicit
form we expand right-hand side of eq. (\ref{Green1}) in power series in quantum
superfields $\phi$, $\bar{\phi}$. As usual, the quadratic part of
expansion of
$\frac{1}{\hbar}S[\bar{\Phi}+\sqrt{\hbar}\bar{\phi},\Phi+\sqrt{\hbar}\phi] $
\begin{eqnarray}
\label{qua}
S_2=\frac{1}{2}\int d^8 z \left(\begin{array}{cc}\phi&\bar{\phi}
\end{array}\right)
\left(\begin{array}{cc}
K_{\Phi\Phi}&K_{\Phi\bar{\Phi}}\\
K_{\Phi\bar{\Phi}}&K_{\bar{\Phi}\bar{\Phi}}
\end{array}\right)
\left(\begin{array}{c}\phi\\
\bar{\phi}
\end{array}
\right)+[\int d^6 z \frac{1}{2}W^{''}\phi^2+h.c.]
\end{eqnarray}
defines the propagators and the higher terms of expansion define the vertices.
Here
$K_{\Phi\bar{\Phi}}=\frac{\partial^2 K(\bar{\Phi},\Phi)}
{\partial\Phi\partial\bar{\Phi}}$,
$K_{\Phi\Phi}=\frac{\partial^2 K(\bar{\Phi},\Phi)}
{\partial\Phi^2}$ etc,  $W^{''}=\frac{d^2 W}{d\Phi^2}$.

%\section{General properties of chiral effective potential}

Let us consider holomorphic effective potential $W_{eff}(\Phi)$.
It was noted by West \cite{West2}
%The mechanism of arising the chiral
%corrections to effective action looks as follows. According to
%non-renormalization theorem (see f.e. \cite{BK0}) all loop corrections to
%effective action can be expressed in form
%of integral over full superspace.
%However, it was pointed by West \cite{West2}
that non-renormalization theorem (see f.e. \cite{BK0})
does not forbid an existence of the finite
chiral corrections to the effective action. The matter is the theories with
chiral superfields admit the loop corrections of the form
%\begin{equation}
%\label{chr}
$\int d^8 z f(\Phi)\big(-\frac{D^2}{4\Box}\big)g(\Phi)=\int d^6 z f(\Phi)g(\Phi)$
%\end{equation}
where $f(\Phi),g(\Phi)$ are some functions of chiral superfield $\Phi$.
We note that this equation %. (\ref{chr})
shows the superfield $\Phi$ is not a constant.
It is easy to prove that the chiral contributions to effective action can be
generated by supergraphs containing massless propagators only.
To find chiral corrections to effective action we put
$\bar{\Phi}=0$ in eqs. (\ref{Green1},\ref{qua}).
{\bf Therefore here and further all derivatives of $K$, $W$ and $\bar{W}$ will
be taken at $\bar{\Phi}=0$.} Under this condition
the action of quantum superfields $\phi,\bar{\phi}$ in external superfield
$\Phi$ looks like
\begin{eqnarray}
\label{qch}
S[\bar{\phi},\phi,\Phi]=\frac{1}{2}\int d^8 z
\left(\begin{array}{cc}\phi&\bar{\phi} \end{array}\right)
\left(\begin{array}{cc}
K_{\Phi\Phi}&K_{\Phi\bar{\Phi}}\\
K_{\Phi\bar{\Phi}}&K_{\bar{\Phi}\bar{\Phi}}
\end{array}\right)
\left(\begin{array}{c}\phi\\
\bar{\phi}
\end{array}
\right)+\int d^6 z \frac{1}{2}W^{''}\phi^2+\ldots
\end{eqnarray}
The dots here denote the terms of third, fourth and higher orders in quantum
superfields.
We  call the theory massless if $W^{''}|_{\Phi=0}=0$.
Further we consider only massless theory.

To calculate the corrections to $W(\Phi)$ we use supergraph technique
(see f.e. \cite{BK0}).
For this purpose one splits the action (\ref{qch}) into sum of free
part and vertices of interaction.
As a free part we take the action $S_0=\int d^8 z \phi\bar{\phi}$.
The corresponding superpropagator is
%\begin{equation}
%\label{pr1}
$G(z_1,z_2) = -\frac{D^2_1\bar{D}^2_2}{16\Box}\delta^8(z_1-z_2)$.
%\end{equation}
And the term $S[\bar{\phi},\phi,\Phi]-S_0$
will be treated as vertices where $S[\bar{\phi},\phi,\Phi]$
is given by eq. (\ref{qch}). Our purpose is to find first leading contribution to
$W_{eff}(\Phi)$. As we will show, chiral loop contributions are began with two
loops. Therefore we keep in eq. (\ref{qch})  only the terms of second, third and fourth
orders in quantum fields.

Non-trivial corrections to chiral potential can arise only
if $2L+1-n_{W^{''}}-n_{V_c}=0$
where $L$ is a number of loops, $n_{W^{''}}$ is a number of vertices
proportional to $W^{''}$, $n_{V_c}$ is that one of vertices of third and
higher orders in quantum superfields,
otherwise corresponding contribution
will either vanish or lead to singularity in infrared linit.
In one-loop
approximation this equation leads to $n_{W^{''}}+n_{V_c}=3$.
However, all supergraphs satisfying this condition have zero contribution.
Therefore first correction to chiral effective potential is two-loop one.
In two-loop approximation this equation has the form $n_{W^{''}}+n_{V_c}=5$.
Since the number of purely chiral (antichiral)
vertices independent of $W^{''}$ in two-loop supergraphs can be equal
to 0, 1 or 2, number of external vertices $W^{''}$ takes values from 3
to 5.

We note that non-trivial contribution to holomorphic
effective potential from any diagram can arise only if number of
$D^2$-factors is more by one than the number of $\bar{D}^2$-factors
(see details in \cite{Buch5}).
The only Green function in the theory is a propagator
$<\phi\bar{\phi}>$. Therefore total number of quantum chiral superfields
$\phi$ corresponding to all vertices
must be equal to that one of antichiral ones $\bar{\phi}$.
As a result we find that the only two-loop supergraph contributing to
chiral effective potential looks like

\vspace*{1mm}

\hspace{4.5cm}
%\vspace*{-1cm}
\unitlength=.6mm
%\linethickness{1.2pt}
%\Lengthunit=1.5cm
%\Linewidth{1.2pt}
%\GRAPH(hsize=3){%\ind(0,-16){Fig.2}%\thicklines
\small
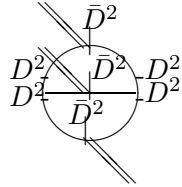
\vspace*{2mm}
\begin{picture}(30,30)
\put(-10,10){\circle{20}}\put(-20,10){\line(1,0){20}}
\put(-10,13){$\bar{D}^2$}
\put(-11,20){$|$}\put(-11,10){$|$}\put(-12,0){$|$}
\put(-15,3){$\bar{D}^2$}
\put(-12,24){$\bar{D}^2$}
\put(-28,13){$D^2$} \put(-28,7){$D^2$} \put(2,13){$D^2$} \put(2,8){$D^2$}
\put(-21,12){-} \put(-21,7){-} \put(0,12){-} \put(0,7){-}
%\Linewidth{0.3pt}
\put(-10,20){\line(-1,1){10}}\put(-11.5,20){\line(-1,1){10}}
\put(-10,0){\line(1,-1){10}}\put(-11.5,0){\line(1,-1){10}}
\put(-10,10){\line(-1,1){10}}\put(-11.5,10){\line(-1,1){10}}
%}
%}
\end{picture}
\normalsize

\vspace*{3mm}

Here internal line is a propagator $<\phi\bar{\phi}>$ depending on background
chiral superfields which has the form
\begin{equation}
\label{propc}
<\phi\bar{\phi}>=-\bar{D}^2_1 D^2_2
\frac{\delta^8(z_1-z_2)}{16 K_{\Phi\bar{\Phi}}(z_1)\Box}.
\end{equation}
We note that the superfield $K_{\Phi\bar{\Phi}}$ is not constant here.
Double external lines are $W''$.

After $D$-algebra transformations and loop integrations
we find that two-loop contribution to holomorphic effective
potential in this model looks like
\begin{equation}
\label{l2c}
W^{(2)}=
\frac{6}{{(16\pi^2)}^2}\zeta(3) \bar{W}^{'''2}
{\Big\{\frac{W''(z)}{K^2_{\Phi\bar{\Phi}}(z)}\Big\}}^3
\end{equation}
One reminds that $\bar{W}^{'''}=\bar{W}^{'''}(\bar{\Phi})|_{\bar{\Phi}=0}$ and
$K_{\Phi\bar{\Phi}}(z)=\frac{\partial ^2 K(\bar{\Phi},\Phi)}
{\partial\Phi\partial\bar{\Phi}}
|_{\bar{\Phi}=0}$ here.
We see that the correction (\ref{l2c}) is finite and does not require
renormalization.

Now we turn to studying of quantum contributions to k\"{a}hlerian effective
potential
%being the superfield analog of standard Coleman - Weinberg potential and
depending only on
superfields $\Phi$, $\bar{\Phi}$ but not on their derivatives.

The one-loop diagrams contributing to kahlerian effective potential are

\hspace{0.5cm}
\unitlength=.6mm
%\thicklines
%\GRAPH(hsize=3){%\ind(50,-30){¨á.1.}
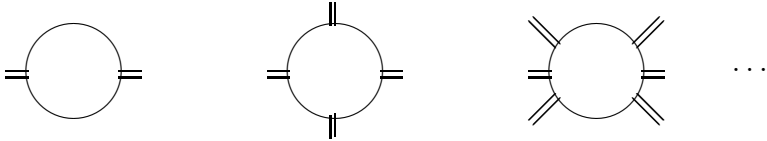
\begin{picture}(20,20)
\put(0,10){\circle{20}}\put(-10,10){\line(-1,0){5}}\put(-10,8.5){\line(-1,0){5}}
\put(10,10){\line(1,0){5}}\put(10,8.5){\line(1,0){5}}
%\ind(24,0){W''}\ind(-26,0){\bar{W''}}
%\put(4.8,0){\GRAPH(hsize=3){
\end{picture}
\hspace{2cm}
\begin{picture}(20,20)
\put(0,10){\circle{20}}\put(-10,10){\line(-1,0){5}}
\put(-10,8.5){\line(-1,0){5}}
\put(10,10){\line(1,0){5}}\put(10,8.5){\line(1,0){5}}
\put(0,20){\line(0,1){5}}\put(-1,20){\line(0,1){5}}
\put(0,0.5){\line(0,-1){5}}
\put(-1,0){\line(0,-1){5}}
%}}
%\ind(-10,20){W''}\ind(-10,-20){W''}
%\ind(-49,0){\bar{W}''}\ind(5,0){\bar{W''}}
%\put(30,0){\ldots}
\end{picture}
%}}}}
\hspace{2cm}
\begin{picture}(30,30)
\put(0,10){\circle{20}}
\put(-10,10){\line(-1,0){5}}\put(-10,8.5){\line(-1,0){5}}
\put(10,10){\line(1,0){5}}\put(10,8.5){\line(1,0){5}}
\put(-9,5){\line(-1,-1){6}}\put(-8,4){\line(-1,-1){6}}
\put(9,5){\line(1,-1){6}}\put(8,4){\line(1,-1){6}}
\put(-9,15){\line(-1,1){6}}\put(-8,16){\line(-1,1){6}}
\put(9,15){\line(1,1){6}}\put(8,16){\line(1,1){6}}
\put(30,10){\ldots}
\end{picture}

\vspace{2mm}

Double external lines correspond to alternating $W''$ and $\bar{W}''$.
Internal lines are $<\phi\bar{\phi}>$-propagators of
the form
$$
G_0\equiv<\phi\bar{\phi}>=-\frac{\bar{D}^2 D^2}{16 K_{\Phi\bar{\Phi}}\Box}
$$
Supergraph of such structure with $2n$
legs represents itself as a ring containing of $n$ links of the following form

\hspace{2cm}
\unitlength=2mm
%\GRAPH(hsize=3){
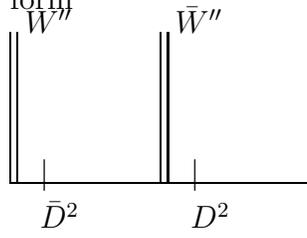
\begin{picture}(40,15)
\put(10,4){\line(1,0){20}}\put(10,4){\line(0,1){10}}\put(10.5,4){\line(0,1){10}}
\put(20,4){\line(0,1){10}}\put(20.5,4){\line(0,1){10}}
%\ind(20,-10){¨á.2}
\put(12,1){$\bar{D}^2$}\put(12,4){$|$}\put(22,1){$D^2$}\put(22,4){$|$}
\put(11,14){$W''$}\put(21,14){$\bar{W}''$}
\end{picture}
%\vspace*{-2mm}

The total contribution of all these diagrams
after $D$-algebra transformations, summarizing, integration over momenta
and subtraction of divergences is equal to
\begin{equation}
\label{k1}
K^{(1)}=-\int d^4 \theta
\frac{1}{32\pi^2}\frac{W''\bar{W}''}{K^2_{\Phi\bar{\Phi}}}%\Big\{
\ln\Big(\frac{W''\bar{W}''}{\mu^2 K^2_{\Phi\bar{\Phi}}}\Big)
%\Big\}
\end{equation}
%It is easy to convince that the present result corresponds to the known
%expression for the Wess-Zumino model where $W''=\frac{1}{2}\lambda\Phi$.

We can find also two-loop correction to k\"{a}hlerian potential.
Since k\"{a}hlerian effective potential depends only on superfields
$\Phi,\bar{\Phi}$ but not on their derivatives supergraphs contributing to
it must include equal number of $D^2$ and $\bar{D}^2$ factors with all
vertices rewritten in the form of an integral over whole superspace.

Components of matrix superpropagator for the case of constant superfields
looks like
\begin{eqnarray}
\label{sol1}
G_{++}&=&\frac{\bar{W}^{''}}{K_{\Phi\bar{\Phi}}\Box+{|W^{''}|}^2}
\frac{\bar{D}^2_1}{4}\delta_{12};\ \
G_{+-}=\frac{1}{K_{\Phi\bar{\Phi}}\Box+{|W^{''}|}^2}
\frac{\bar{D}^2_1 D^2_2}{16}\delta_{12}\nonumber\\
G_{-+}&=&\frac{1}{K_{\Phi\bar{\Phi}}\Box+{|W^{''}|}^2}
\frac{D^2_1\bar{D}^2_2}{16}\delta_{12};\ \
G_{--}=\frac{W^{''}}{K_{\Phi\bar{\Phi}}\Box+{|W^{''}|}^2}
\frac{D^2_1}{4}\delta_{12}
\end{eqnarray}
The all diagrams with equal number of $D^2$ and $\bar{D}^2$-factors are
given in this picture.

%\vspace{2mm}

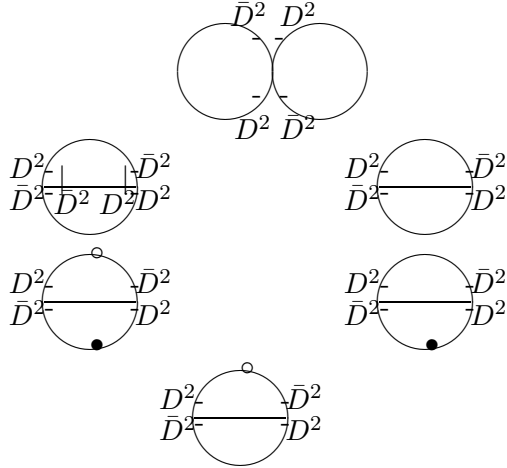
\small

\hspace{6cm}
\unitlength=.6mm
\begin{picture}(30,20)
%\Lengthunit=1.2cm
%\GRAPH(hsize=3){%\ind(10,-15){Fig.3.1}
%\ind(-7,0){G_{--}} \ind(30,0){G_{--}}
\put(5,0){\circle{20}}\put(26,0){\circle{20}}
\put(11,-7){-}\put(17,-7){-}
\put(11,6){-}\put(16,6){-}
\put(7,-15){$D^2$}\put(17,-15){$\bar{D}^2$}
\put(6,10){$\bar{D}^2$}\put(17,10){$D^2$}
%}
\end{picture}

\vspace*{3mm}

\hspace{4.5cm}
%\Lengthunit=1.2cm
%\GRAPH(hsize=3){%\ind(0,-15){Fig.3.2}
\begin{picture}(20,20)
\put(0,0){\circle{20}}\put(-10,0){\line(1,0){20}}
%\put(-8,0){$|$}\ind(-6,2){\bar{D}^2}
%\ind(6,0){|}\ind(5,2){D^2}
\put(-18,2){$D^2$} \put(-18,-5){$\bar{D}^2$} \put(10,2){$\bar{D}^2$}
\put(10,-5){$D^2$}
\put(-10,2){-} \put(-10,-3){-} \put(9,2){-} \put(9,-3){-}
\put(-7,0){$|$}\put(7,0){$|$}
\put(-8,-6){$\bar{D}^2$}\put(2,-6){$D^2$}
\end{picture}
\hspace*{3cm}
\begin{picture}(20,20)
\put(0,0){\circle{20}}\put(-10,0){\line(1,0){20}}
%\put(-8,0){$|$}\ind(-6,2){\bar{D}^2}
%\ind(6,0){|}\ind(5,2){D^2}
\put(-18,2){$D^2$} \put(-18,-5){$\bar{D}^2$} \put(10,2){$\bar{D}^2$}
\put(10,-5){$D^2$}
\put(-10,2){-} \put(-10,-3){-} \put(9,2){-} \put(9,-3){-}
\end{picture}

\vspace*{3mm}

\hspace{4.5cm}
\begin{picture}(20,20)
\put(0,0){\circle{20}}\put(-10,0){\line(1,0){20}}
%\put(-8,0){$|$}\ind(-6,2){\bar{D}^2}
%\ind(6,0){|}\ind(5,2){D^2}
\put(-18,2){$D^2$} \put(-18,-5){$\bar{D}^2$} \put(10,2){$\bar{D}^2$}
\put(10,-5){$D^2$}\put(0,9.5){$\circ$}\put(0,-11){$\bullet$}
\put(-10,2){-} \put(-10,-3){-} \put(9,2){-} \put(9,-3){-}
\end{picture}
\hspace*{3cm}
\begin{picture}(20,20)
\put(0,0){\circle{20}}\put(-10,0){\line(1,0){20}}
%\put(-8,0){$|$}\ind(-6,2){\bar{D}^2}
%\ind(6,0){|}\ind(5,2){D^2}
\put(-18,2){$D^2$} \put(-18,-5){$\bar{D}^2$} \put(10,2){$\bar{D}^2$}
\put(10,-5){$D^2$}\put(0,-11){$\bullet$}
\put(-10,2){-} \put(-10,-3){-} \put(9,2){-} \put(9,-3){-}
\end{picture}
%}
%}

\vspace*{3mm}

\hspace{6.5cm}
\begin{picture}(20,20)
\put(0,0){\circle{20}}\put(-10,0){\line(1,0){20}}
%\put(-8,0){$|$}\ind(-6,2){\bar{D}^2}
%\ind(6,0){|}\ind(5,2){D^2}
\put(-18,2){$D^2$} \put(-18,-5){$\bar{D}^2$} \put(10,2){$\bar{D}^2$}
\put(10,-5){$D^2$}\put(0,9.5){$\circ$}
\put(-10,2){-} \put(-10,-3){-} \put(9,2){-} \put(9,-3){-}
\end{picture}

%\vspace*{3mm}
\newpage

Here

\vspace*{2mm}

\hspace*{2cm}
%\GRAPH(hsize=3){
\begin{picture}(40,20)
\put(7,20){\line(1,0){20}}\put(37,20)
{$=\frac{1}{K_{\Phi\bar{\Phi}}\Box+{|W^{''}|}^2}$}
\put(85,20){\line(1,0){20}}\put(95,20){$\circ$}
\put(109,20){$=\frac{W^{''}}{K_{\Phi\bar{\Phi}}\Box+{|W^{''}|}^2}$}
\put(6,7){\line(1,0){20}}\put(16,7){$\bullet$}
\put(35,7){$=\frac{\bar{W}^{''}}{K_{\Phi\bar{\Phi}}\Box+{|W^{''}|}^2}$}
\end{picture}

\normalsize
\vspace*{2mm}

Two-loop correction to k\"{a}hlerian potential is a sum of all
these contributions.  After $D$-algebra transformations, loop integrations
and subtraction of one-loop and two-loop divergences
it is equal to
\begin{eqnarray}
\label{2loop}
K^{(2)}&=&\frac{1}{{(16\pi^2)}^2}
\Big(\frac{{|W''|}^2}{{K_{\Phi\bar{\Phi}}}^4}(-\frac{1}{4}\ln^2
\frac{{|W''|}^2}{\mu^2 K_{\Phi\bar{\Phi}}}
+\frac{3-\gamma}{2}\ln\frac{{|W''|}^2}{\mu^2 K_{\Phi\bar{\Phi}}}
+\frac{3}{2}(\gamma-1)+\nonumber\\&+&\frac{1}{4}(\gamma^2+\zeta(2)))\Big)
\Big\{
\frac{1}{6}\big({K_{\Phi\bar{\Phi}}}^3 {|W'''|}^2+(\bar{W}''W'''
{K_{\Phi\bar{\Phi}}}^2
K_{\Phi\Phi\bar{\Phi}}+h.c.)+\nonumber\\&+&
W''\bar{W''} K_{\Phi\bar{\Phi}} K_{\Phi\Phi\bar{\Phi}}
K_{\Phi\bar{\Phi}\bar{\Phi}}+\\&+&
{K_{\Phi\bar{\Phi}}}^2 {|W''|}^2 |K_{\Phi\Phi\bar{\Phi}}|^2
\big)+
\frac{1}{2}{K_{\Phi\bar{\Phi}}}^2 {|W''|}^2K_{\Phi\Phi\bar{\Phi}}
K_{\Phi\bar{\Phi}\bar{\Phi}}
\Big\}+\frac{1}{{(16\pi^2)}^2}\times\nonumber\\&\times&
\frac{{|W''|}^4}{{K_{\Phi\bar{\Phi}}}^4}(\gamma-1+
\ln\frac{{|W''|}^2}{\mu^2 K_{\Phi\bar{\Phi}}})^2
\big({K_{\Phi\bar{\Phi}}}^2K_{\Phi\Phi\bar{\Phi}}K_{\Phi\bar{\Phi}\bar{\Phi}}+
{K_{\Phi\bar{\Phi}}}^2 K_{\Phi\Phi\bar{\Phi}\bar{\Phi}}\big)\nonumber
\end{eqnarray}
The value of normalization point $\mu$ can be fixed by imposing of suitable
normalization condition.

%\section{Summary}
To conclude, we have solved the problem of calculating leading holomorphic
correction to superfield effective action in general chiral superfield model
(1) with arbitrary potentials $K(\bar{\Phi},\Phi)$ and $W(\Phi)$.
The result has the universal form (\ref{l2c}) and it is finite independently
if the functions $K(\bar{\Phi},\Phi)$, $W(\Phi)$ correspond to
renormalizable theory or no.
We have also calculated one-loop and two-loop contributions to k\"{a}hlerian
effective potential.
We note that results (\ref{l2c},\ref{k1},\ref{2loop}) reproduces
the known results for Wess-Zumino model
at $W=-\frac{\lambda}{3!}\Phi^3$, $K=\bar{\Phi}\Phi$
\cite{West2,Buch3}.

{\bf Acknowledgements.} %Author is grateful to M. Cvetic and S.M. Kuzenko
%for discussions.
The work was carried out under
partial support of INTAS, project INTAS--96--0308; RFBR -- DFG, project No.
99-02-04022; RFBR, project No. 99-02-16617; grant center of
St. Peterburg University, project No. 97--6.2--34.

\end{document}